# Discussion on the Physics Problem of a Boat Crossing a River

Kyle Kou Yuchang[1], Simon Meng Zimin[1], Paul Zhang Yixing[2]

[1]Wuhan Britain-China Senior High School, No.10 Gutian Ce Rd., Qiaokou District, Wuhan, Hubei, P.R.China

[2]Physics Department, Wuhan Britain-China Senior High School, No.10 Gutian Ce Rd., Qiaokou District, Wuhan, Hubei, P.R.China

**Abstract**

This study addresses the boat river-crossing problem under non-uniform flow velocities by constructing three models: constant flow (Model 1), linear distribution (Model 2), and even-power function distribution (Model 3, adjustable via parameter $n$ ). By using the vector addition, combined with the solutions of calculus and differential equations, the analytical expression of the ship's spatial trajectory under a fixed heading angle relative to the water flow is derived. For the shortest-time control problem, the Lagrange multiplier method is introduced to construct a constrained optimization model, and the analytical solution of the optimal heading angle that satisfies the boundary condition of reaching the direct opposite bank is solved. The research results provide theoretical support for the path planning of inland ship intelligent navigation systems, and the proposed multi-model analysis framework can effectively simulate the complex flow velocity distribution scenarios of real rivers.

**Keywords:** Kinematics; River crossing problem; Non-uniform flow; Minimum time optimization; Lagrange multiplier; Physics education

## 1. Introduction

The problem of a boat crossing a river has long been a basic example in classical physics. It explains key ideas in kinematics, like relative motion, vector decomposition and trajectory optimization. Almost all of the existing physics papers and problem analyses only assume that the river's flow speed is uniform. They focus on two typical situations: making the boat's speed perpendicular to the riverbank to reduce the crossing time as much as possible, or changing the boat's direction to offset the uniform water flow and make sure the boat reaches the bank directly opposite. The condition in these studies and physics problems is that the flow velocity of the water (in terms of both magnitude and direction) remains constant and is independent of the water's position. Although these idealized models have research value, they ignore one characteristic of real rivers: the middle of most rivers is deep while the banks are shallow. In the middle of a river, the cross-sectional area of water flow is larger, and the constraint on the water flow is relatively smaller, allowing the water to maintain a higher flow velocity. On both sides of the river, however, the cross-sectional area of water flow is relatively smaller, and the constraint on the water flow is greater, resulting in a slower flow velocity [1]. This discrepancy between theoretical assumptions and actual river environments has prompted us to conduct a more detailed exploration of boat motion in non-uniform flow fields. This study fills this gap by constructing three representative flow velocity models, in which the river velocity is defined as a function of the distance from the riverbank. Specifically, we consider the traditional constant flow velocity, as well as flow velocity distribution models where the velocity increases with the lateral distance from the riverbank in the form of linear and even-power functions. Each model captures different degrees of velocity gradients across the river width. The progressive models are designed for physics education, enabling students to explore concepts like vector addition and the applications of calculus through real-world river-crossing examples.

## 2. Theoretical Methodology

In the model, it is assumed that the boat's velocity relative to the water flow is constant at $v$, and the direction of the river flow remains unchanged, flowing downstream. A coordinate system is established with the starting point as the origin, where the x-axis points in the direction of the downstream river flow, and the y-axis points toward the opposite bank, perpendicular to the riverbank. The coordinates of the destination that the boat needs to reach are $(0, L)$, where $L$ represents the width of the river. The angle $\theta$ is the angle between the direction of the boat's velocity and the negative direction of the x-axis. See Figure 1, which shows the coordinate system and velocity decomposition for boat crossing problem. The boat's velocity relative to water is decomposed into components parallel and perpendicular to the riverbank, with angle $\theta$ measured from the negative x-axis. The addition of vectors follows the parallelogram law.

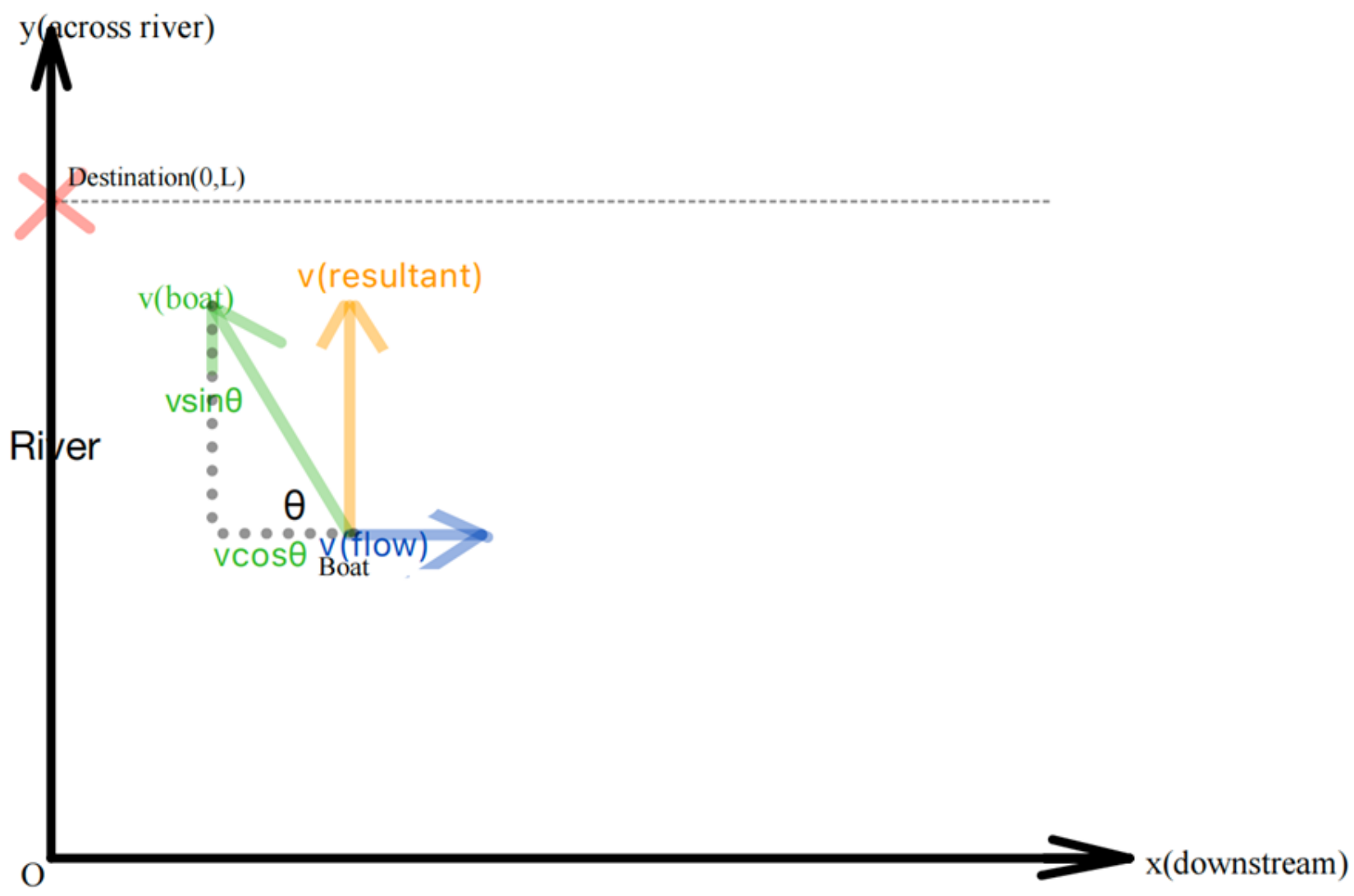

Figure 1: Vector addition of velocity and angle $\theta$.

Symbols and their meanings are shown in Table 1.

| Symbol | Physical Meaning | Unit |
| --- | --- | --- |
| $\theta$ | The angle between the velocity of the boat relative to the water and the negative direction of the x-axis | rad |
| $v_{\text{boat to water}}$ | The magnitude of the velocity of the boat relative to the water | m/s |
| $v(y)$ | The magnitude of the water flow velocity at position $y$ | m/s |
| $L$ | The width of the river | m |
| $T$ | The time to cross the river | s |

Table 1: Symbols, Physical Meanings, and Units Related to Boat Motion in a River.

In each model, I will discuss three questions: 1. When the boat's direction relative to the river's flow velocity remains unchanged, what value should the angle $\theta$ take to ensure the boat reaches the destination $(0, L)$ exactly? 2. What is the function of the path taken by the boat in this case? 3. If the angle $\theta$ can be adjusted, how to minimize the total time for the boat to reach the destination?

## 2.1 Model 1: Constant Flow

### 2.1.1 Description of Model One

In this model, the velocity of the water flow is constant at $kv$ and is independent of the distance to the riverbank.

### 2.1.2 Q1 & Q2

The key to reaching the direct opposite bank is that the component of the boat's resultant velocity relative to the bank, along the direction parallel to the riverbank (and parallel to the water flow), must be zero. Otherwise, the boat will be carried off course by the water flow, resulting in a displacement along the riverbank.

$$cos\theta = \frac{kv}{v} = k \qquad (1)$$

When $k > 1$, the maximum upstream component of the boat's velocity is $v$, which is less than $kv$ (the river's flow velocity). As a result, the resultant velocity still has a positive component along the riverbank (in the downstream direction of the water flow), making it impossible for the boat to reach the direct opposite bank. It is only possible for the boat to reach the direct opposite bank when $k < 1$.

The function of the path is a straight line

$$x = 0 \qquad (2)$$

See Figure 2 for the trajectory. The boat follows a straight path along the y-axis when the heading angle is optimally chosen to counteract the uniform water flow.

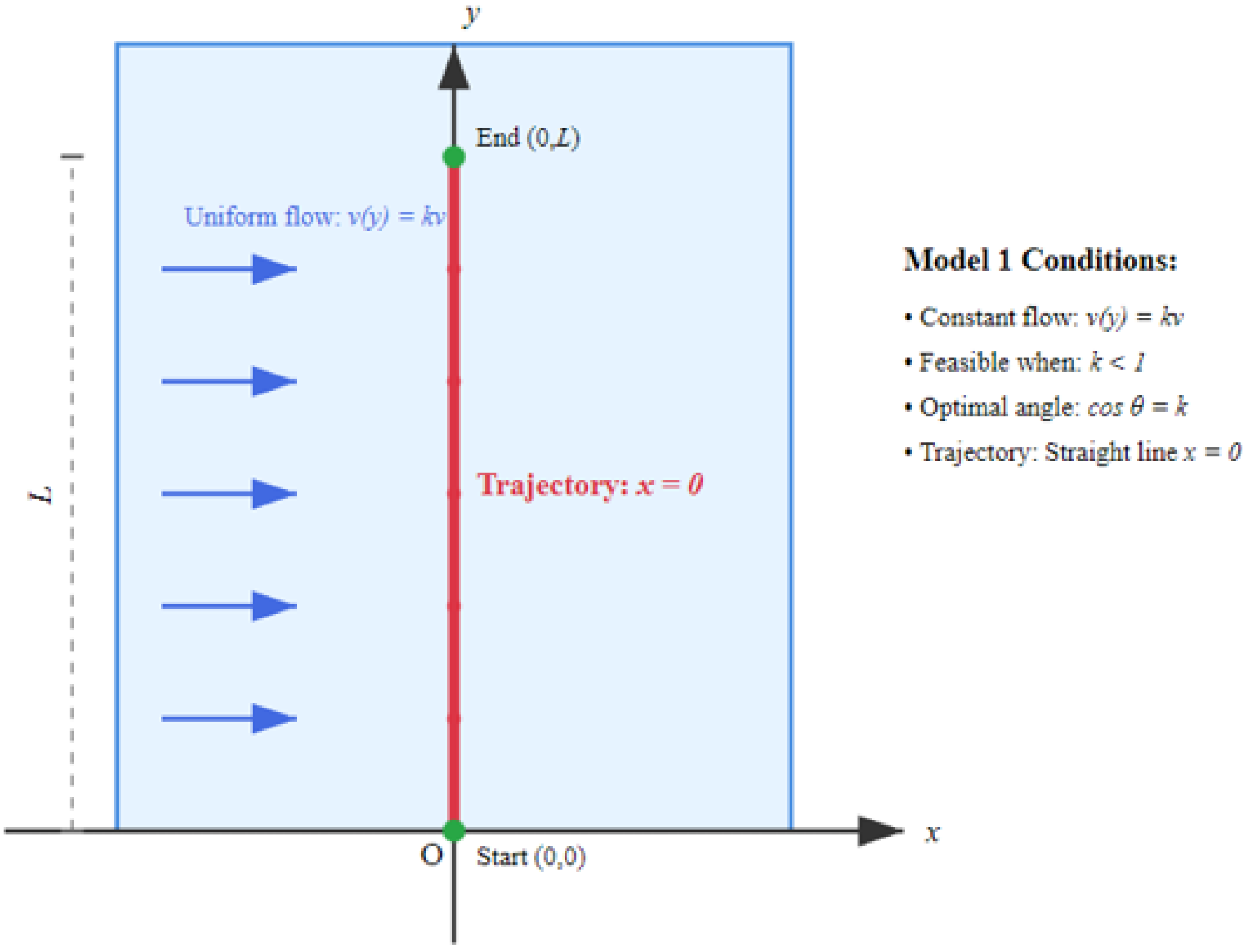

Figure 2: Trajectory of Model 1.

### 2.1.3 Q3

Under the prerequisite that "the boat can reach the direct opposite bank" $k < 1$, the river-crossing time should be minimized by adjusting the angle. The core logic is that the river-crossing time is determined by the component of the resultant velocity perpendicular to the riverbank $v_{boat\ to\ water,y}$, so this component needs to be maximized. Through Pythagorean theorem

$$v_{boat\ to\ water,y} = \sqrt{v^2 - v_{boat\ to\ water,x}{}^2} = v\sqrt{1-k^2} \qquad (3)$$

The river-crossing time is

$$T_1 = \frac{L}{v\sqrt{1-k^2}} \qquad (4)$$

The trajectory is also shown in Figure 1.

## 2.2 Model 2: Linear Distribution

### 2.2.1 Description of the Model

In the second model, the function of the water flow is

$$v(y) = kv\left(1 - \frac{|2y - L|}{L}\right) \qquad (5)$$

and the function graph is axisymmetric about the central axis of the river. This is a triangular distribution. $kv$ represents the velocity of the water flow in the center of the river, which is k times the boat's velocity relative to the water. The water flow velocity is the highest at this location.

Boat’s velocity relative to water:

x-component

$$v_{boat\ to\ water,x} = -v\cos\theta \qquad (6)$$

y-component

$$v_y = v\sin\theta \qquad (7)$$

Boat's absolute velocity (relative to ground):

$$v_x = v(y) + v_{boat\ to\ water,x} = kv\left(1 - \frac{|2y - L|}{L}\right) - v\cos\theta \qquad (8)$$

y-component does not change.

For the boat to reach the point directly opposite, the net x-displacement must be zero.

Time to cross the river:

$$t = \frac{L}{v\sin\theta} \qquad (9)$$

Net x-displacement:

$$\Delta x = \int_0^t v_x dt = \int_0^L v_x \cdot \frac{dy}{v_y} = 0 \qquad (10)$$

### 2.2.2 Q1 & Q2

Interval 1: $y \in [0, \frac{L}{2}]$

Simplify Equation (5):

$$v(y) = kv\left(1 - \frac{L-2y}{L}\right) = \frac{2kvy}{L} \qquad (11)$$

Thus,

$$v_x = \frac{2kvy}{L} - vcos\theta \qquad (12)$$

Integrate:

$$\int_0^{L/2} \left(\frac{2kvy}{L} - vcos\theta\right) dy = v\left(\frac{kL}{4} - \frac{Lcos\theta}{2}\right) \qquad (13)$$

Interval 2: $y \in (\frac{L}{2}, L]$

By the same logic,

$$\int_{L/2}^{L} \left(2kv - \frac{2kvy}{L} - vcos\theta\right) dy = vL\left(\frac{k}{4} - \frac{cos\theta}{2}\right) \qquad (14)$$

Sum the two integrals and set to zero (for $\Delta x = 0$):

$$vL\left(\frac{k}{4} - \frac{cos\theta}{2} + \frac{k}{4} - \frac{cos\theta}{2}\right) = 0 \qquad (15)$$

$$\theta = \arccos\left(\frac{k}{2}\right) \qquad (16)$$

In this case $k$ must be smaller than 2 since the cosine function must be less than 1 here. If $k \geq 2$, flow velocity exceeds the boat's capability.

Now we proceed to derive the motion trajectory of the boat.

$$sin\theta = \frac{\sqrt{4-k^2}}{2} \qquad (17)$$

$$v_y = \frac{\sqrt{4-k^2}}{2} v \qquad (18)$$

Interval 1: $y \in [0, \frac{L}{2}]$

From Equation (12),

$$v_x = \frac{2kvy}{L} - v \cdot \frac{k}{2} = v\left(\frac{2ky}{L} - \frac{k}{2}\right) \qquad (19)$$

$$dx = v_x dt = v\left(\frac{2ky}{L} - \frac{k}{2}\right) \cdot \frac{2dy}{\sqrt{4-k^2}v} = \frac{2}{\sqrt{4-k^2}}\left(\frac{2ky}{L} - \frac{k}{2}\right) dy \qquad (20)$$

$$x(y) = \frac{2}{\sqrt{4-k^2}} \int_0^y \left(\frac{2ky}{L} - \frac{k}{2}\right) dt = \frac{2k}{\sqrt{4-k^2}} \left(\frac{y^2}{L} - \frac{y}{2}\right) \qquad (21)$$

Interval 2: $y \in (\frac{L}{2}, L]$

After integrating from $\frac{L}{2}$ to $y$, the result can be obtained by the same logic that

$$x(y) = \frac{2k}{\sqrt{4-k^2}} \left(-\frac{y^2}{L} + \frac{3y-L}{2}\right) \qquad (22)$$

See Figure 3, which shows the shapes of Function (21) and Function (22).

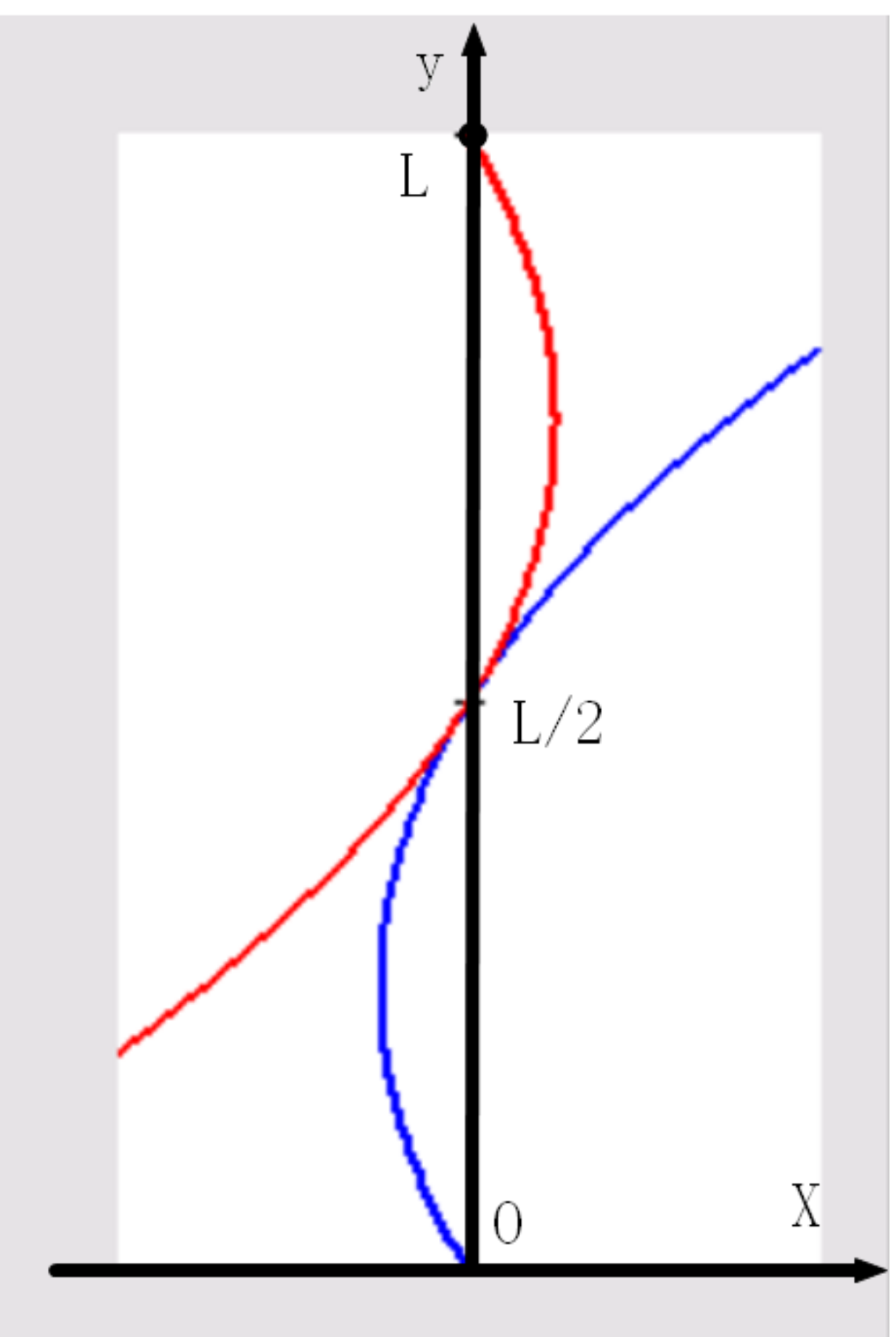

Figure 3: The shape of Function (21) and Function (22).

Considering that Function (21) and Function (22) are two interval functions, the parts outside the intervals are removed, resulting in Figure 4, which shows the boat's trajectory.

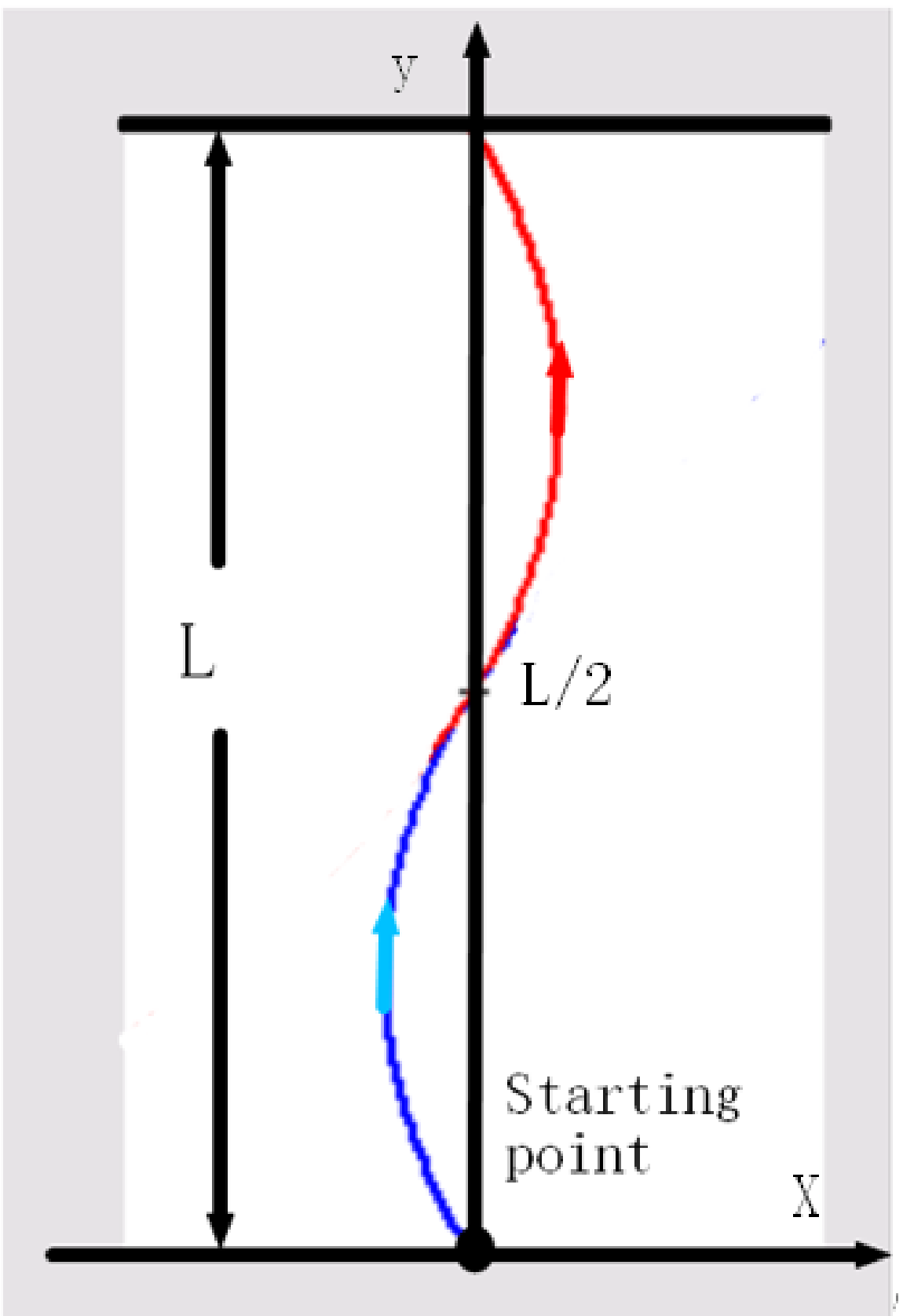

Figure 4: The Trajectory of Model 2 (linear Distribution).

**2.2.3 Q3**

The angle between the direction in which the boat travels and the straight line along which the water flows can be changed arbitrarily, so the angle $\theta$ is denoted as $\theta(y)$

$x$-component (downstream positive): $v_x(y) = -vcos\theta(y) + v(y)$ (the ship's $x$-velocity relative to water points to the negative $x$-direction, superposed with the positive flow velocity).

$y$-component (perpendicular to the riverbank positive): $v_y(y) = vsin\theta(y)$ (unchanged, as the perpendicular direction is unaffected by the flow's $x$-component).

The river-crossing time $T$ is the integral of the reciprocal of the $y$-component velocity over $y \in [0, L]$:

$$T = \int_0^L \frac{dy}{v_y(y)} = \frac{1}{v}\int_0^L \frac{dy}{sin\theta(y)} \tag{23}$$

To reach the direct opposite bank, the total x-displacement must be zero:

$$\int_0^T v_x(t)dt = 0 \qquad (24)$$

Substituting $dt = \frac{dy}{v_y(y)}$ to change the integration variable to $y$, we derive:

$$\int_0^L \left[ -cot\theta(y) + \frac{k\left(1 - \frac{|2y-L|}{L}\right)}{sin\theta(y)} \right] dy = 0 \qquad (25)$$

To minimize $T$ under the above constraint, construct the Lagrangian functional [2]:

$$F = \int_0^L \frac{1}{vsin\theta(y)} + \lambda \left( -cot\theta(y) + \frac{k\left(1 - \frac{|2y-L|}{L}\right)}{sin\theta(y)} \right) dy \qquad (26)$$

where $\lambda$ is the Lagrange multiplier.

Taking the variation of $\theta(y)$ (setting $\delta F = 0$), we compute the partial derivative of the integrand with respect to $\theta$ and set it to zero. Let the integrand be:

$$f = \frac{1}{vsin\theta(y)} + \lambda \left( -cot\theta(y) + \frac{k\left(1 - \frac{|2y-L|}{L}\right)}{sin\theta(y)} \right) \qquad (27)$$

Take the partial derivative of $f$ with respect to $\theta$ and simplify. This leads to:

$$cos\theta(y) = \frac{-\beta}{1 - \beta k(1 - \frac{|2y-L|}{L})} \qquad (28)$$

where $\beta = \lambda v$ is a scaled Lagrange multiplier.

If the ship's heading is tilted upstream $(cos\theta > 0)$, $\beta < 0$ (ensuring the numerator $-\beta > 0$ and the denominator positive, consistent with $cos\theta > 0$).

Using

$$\theta(y) = \theta(L-y) \qquad (29)$$

$$\int_0^1 \frac{-\beta k^2\tau^2 + k\tau + \beta}{\sqrt{(1-\beta k\tau)^2 - \beta^2}} d\tau = 0 \qquad (30)$$

where

$$\tau = \frac{2y}{L} \qquad (\tau \in [0,1]) \qquad (31)$$

This integral is solved numerically to find $\beta$ (dependent on $k$).

For example, consider a special case $k = 1$. One of the authors of this paper, Kyle Kou Yuchang, discussed this special case of the minimum time in his IB Mathematics Internal Assessment (IA) [3].

$$\beta = -0.6225 \qquad (32)$$

$$T = \frac{1.152L}{v} \qquad (33)$$

#### 2.2.4 Results

When the water flow velocity has a linear growth relationship with the distance to the riverbank, the boat can reach the direct opposite bank when the angle between the boat and the river (the opposite of the flow direction) is

$$\theta = \arccos\left(\frac{k}{2}\right) \qquad (34)$$

In this situation, the trajectory of the boat is

$$x(y) = \begin{cases} \frac{2k}{\sqrt{4-k^2}}\left(\frac{y^2}{L} - \frac{y}{2}\right), & 0 \le y \le \frac{L}{2} \qquad (21) \\ \frac{2k}{\sqrt{4-k^2}}\left(-\frac{y^2}{L} + \frac{3y-L}{2}\right), & \frac{L}{2} < y \le L \qquad (22) \end{cases}$$

To minimize the time taken for the boat to travel, solve the calculus using Equation (30) to find $\beta$, and then substitute it into Equation (28).

In summary, minimizing the crossing time requires using variational calculus (Lagrangian multiplier method) to derive the angle $\theta(y)$, then solving for the multiplier $\beta$ numerically, and finally computing the time via integration.

## 2.3 Model 3: Power-Law Distribution

#### 2.3.1 Description of the Model

In the second model, the function of the water flow is

$$v(y) = \text{kv}\left[1 - (\frac{2y - L}{L})^{2n}\right] \qquad (35)$$

where $n$ is a positive integer. Such a function is obtained by translating an even function, which ensures symmetry. The function graph of this velocity with respect to the distance from the riverbank where the starting point is located is shown in Figure 5.

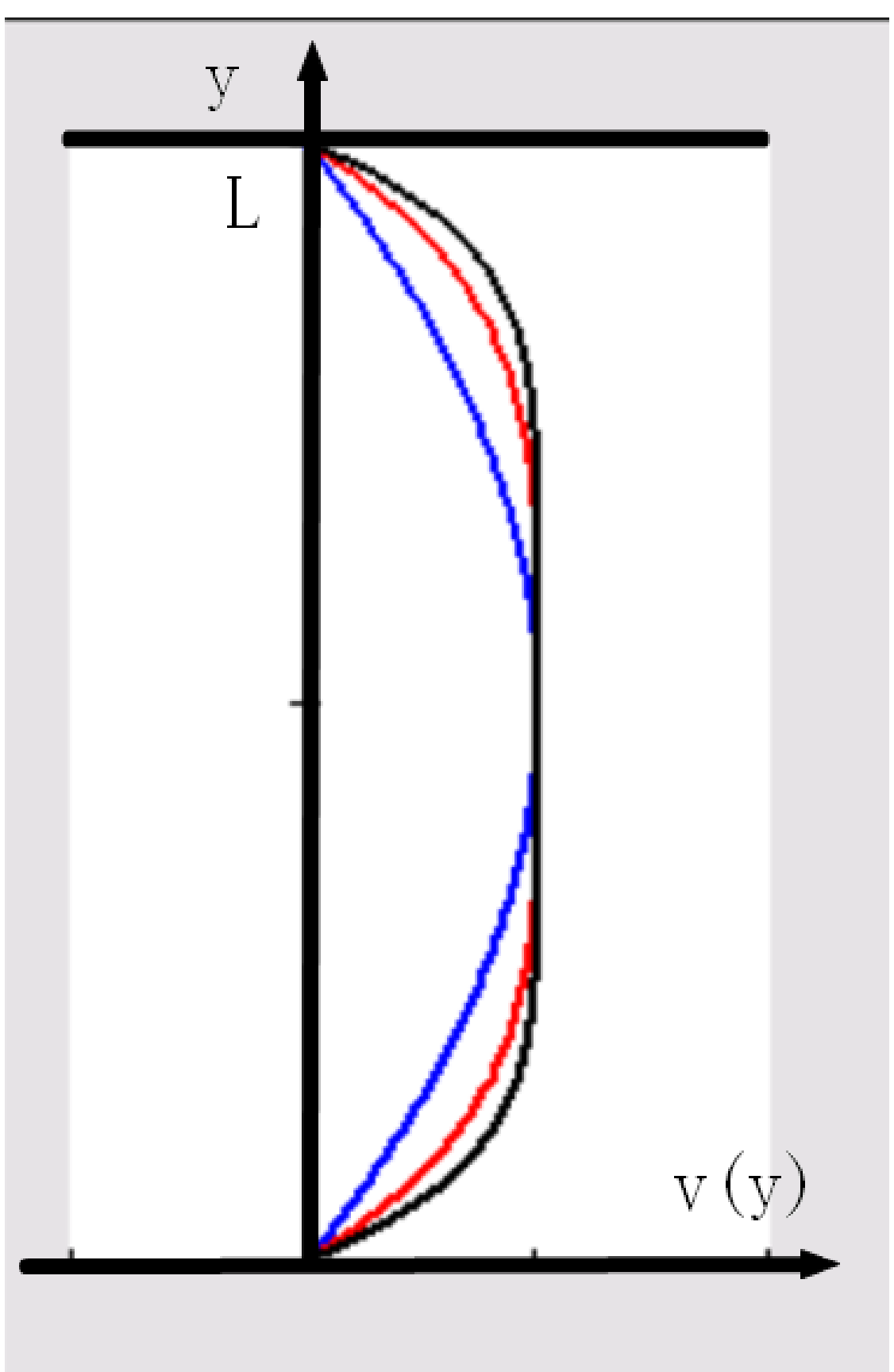

Figure 5: Shape of Function (35), which shows the y position of the water against the velocity.

Boat's velocity relative to water:

x-component

$$v_{boat\ to\ water,x} = -v\cos\theta \qquad (36)$$

y-component

$$v_y = v\sin\theta \qquad (37)$$

Boat's absolute velocity (relative to ground):

$$v_x = v(y) - vcos\theta \qquad (38)$$

Using the same method as in the second model, set the displacement in the x-direction to zero,

$$\int_0^L [v(y) - vcos\theta]\frac{dy}{vsin\theta} = 0 \qquad (39)$$

### 2.3.2 Q1 & Q2

Using substitution to solve the integration in (38), we derive:

$$\frac{L}{2}\left[k\left(2 - \frac{2}{2n+1}\right) - 2cos\theta\right] = 0 \qquad (40)$$

$$\theta = \arccos\left(\frac{2kn}{2n+1}\right) \qquad (41)$$

Now, in order to find the motion trajectory of the boat, use Equation (38)

$$\int_0^x dx = \frac{1}{sin\theta}\int_0^t \left\{k\left[1 - \left(\frac{2t-L}{L}\right)^{2n}\right] - cos\theta\right\}dt \qquad (42)$$

$$x(y) = \frac{kL}{2(2n+1)sin\theta}\left[\frac{2y-L}{L} - \left(\frac{2y-L}{L}\right)^{2n+1}\right] \qquad (43)$$

where $n$ is a positive integer. Function (42) is centrally symmetric about $(\frac{L}{2}, 0)$, two symmetric extreme points exist. In the following, different positive integer values of n will be substituted into the function to find the shape of the trajectory. See Figure 6.

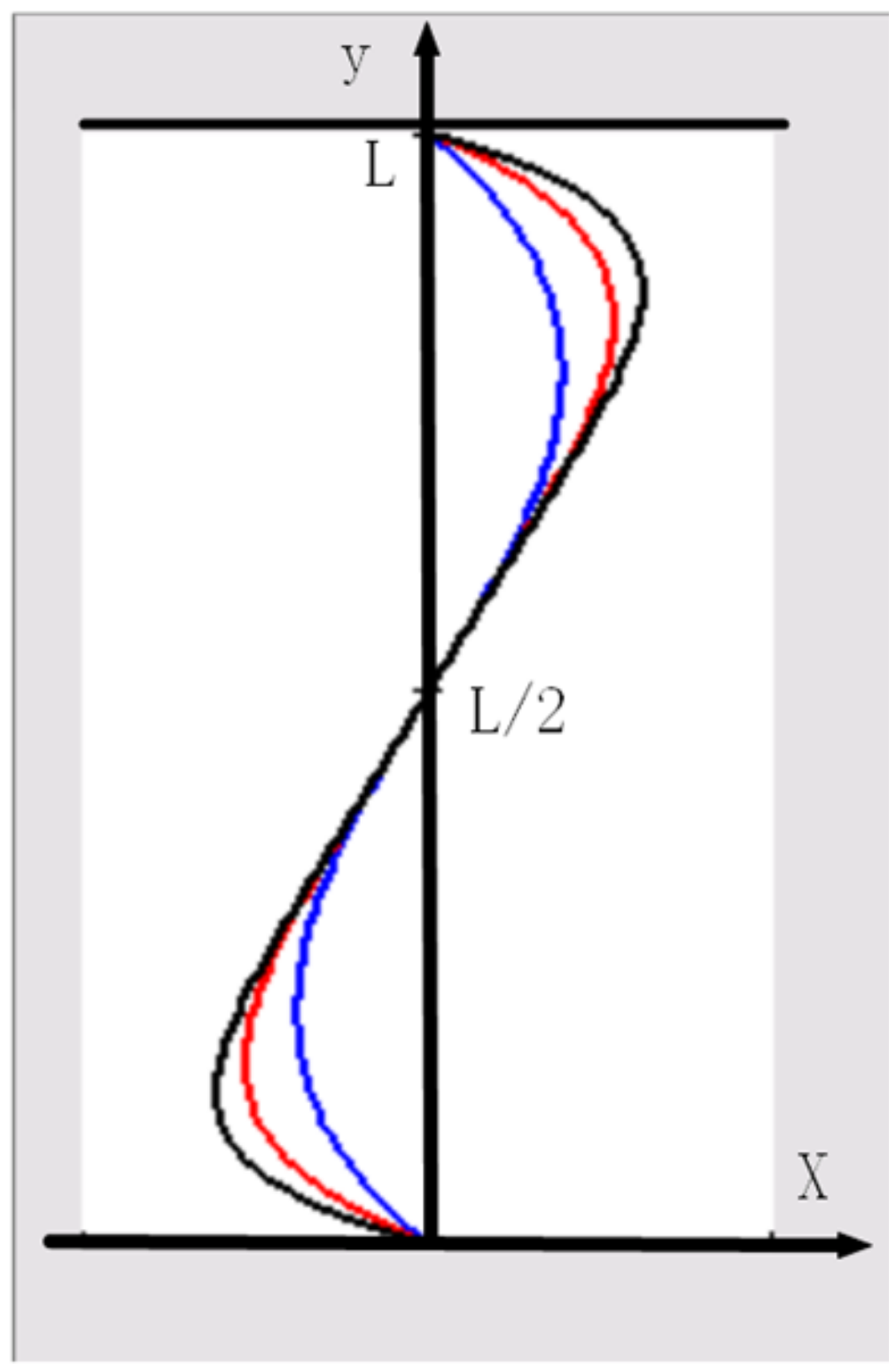

Figure 6: Shapes of the trajectory. (The blue line represents the case when $n=1$; The red line represents the case when $n=2$; The black line represents the case when $n=3$.)

Here is the explanation of the shape of the trajectory. When $n$ increases, the term $(\frac{2y-L}{L})^{2n}$ (an even-power term) decays extremely rapidly. Even at positions close to the banks (where y deviates from the river center $\frac{L}{2}$), this power term remains small, causing $v(y)$ to be close to $kv$ (the maximum flow velocity at the river center), as shown in Figure 5. As $n$ increases, the flow velocity becomes closer to being uniform $kv$. Therefore, the ship requires a larger velocity component in the -x direction. Therefore, the maximum displacement of the ship in the x-direction will increase. Meanwhile, the extreme point will move closer to the riverbank, because the water flow velocity near the bank becomes closer to the high velocity in the middle of the river.

To verify this conclusion, select specific values, such as $k=\frac{3}{4}$ and $n=1$. This special case is exactly a problem from the book *Physics problems for aspiring physical scientists and engineers: With hints and full solutions* [4]. Substitute the values into the formula derived from the theoretical deduction, and we derive:

$$\theta = \arccos\left(\frac{1}{2}\right) = 60° \qquad (44)$$

It matches the answer in the book.

### 2.3.3 Q3

We can obtain the following results using the same method as described in Section **2.2.3 Q3**.

$$cos\theta = \frac{\beta}{\beta k\left[1 - \left(\frac{2y - L}{L}\right)^{2n}\right] + 1} \qquad (45)$$

and

$$\int_{-1}^{1}\left[k(1 - z^{2n}) - \frac{\beta}{k(1 - z^{2n}) + \beta - 1}\right]\frac{dz}{\sqrt{1 - (\frac{\beta}{k(1 - z^{2n}) + \beta - 1})^2}} = 0 \qquad (46)$$

where

$$z = \frac{2y - L}{L} \qquad (47)$$

to simplify the 2n-th order term. This integral is solved numerically to find $\beta$.

Once $\beta$ is found, substitute it back into $cos\theta(y)$, then compute $\theta(y) = arccos(cos\theta(y))$. This $\theta(y)$ minimizes the crossing time while satisfying zero downstream displacement.

### 2.3.4. Results

When the water flow velocity follows the even-power function distribution (Model 3), the boat can reach the direct opposite bank when the fixed angle between its heading and the river flow direction satisfies

$\theta = \arccos\left(\frac{2kn}{2n+1}\right) \qquad (41),$

where $k$ represents the ratio of the river's maximum flow velocity (at the center) to the boat's velocity relative to water, and $n$ (a positive integer) determines the non-linearity of the flow velocity distribution.

In this scenario, the trajectory of the boat is expressed as a piecewise-continuous function:

$$x(y)=\frac{kL}{2(2n+1)sin\theta}\left[\frac{2y-L}{L}-\left(\frac{2y-L}{L}\right)^{2n+1}\right] \qquad (43)$$

This trajectory is centrally symmetric about the point $\left(\frac{L}{2},0\right)$ and has two symmetric extreme points; its curvature and offset change with $n$ (e.g., smaller $n$ leads to more obvious offset, larger $n$ makes the curve flatter), as shown in Figure 4.

To minimize the boat's crossing time, the heading angle $\theta$ needs to be adjusted dynamically with the lateral position $y$ (denoted as $\theta(y)$). Using the same variational calculus (Lagrangian multiplier method) as in Section 2.2.3, we first solve the scaled Lagrange multiplier $\beta$ numerically via the integral equation (related to $z=\frac{2y-L}{L}$), then substitute $\beta$ into the expression of $\cos\theta(y)$ (derived in 2.3.3) to obtain the optimal $\theta(y)$. Finally, substitute $\theta(y)$ into the integral of $\frac{dy}{v\sin\theta(y)}$ to compute the minimum crossing time.

# 3. Discussion

This study builds three different flow velocity distribution models and analyzes the movement rules of a small boat in a non-uniform flow field in a systematic way. Through comparative analysis, we have found the following important rules:

## 3.1 Connection and Unity Between Models

The three models actually represent a progressive relationship in the complexity of flow velocity distribution. The constant flow velocity model (Model 1) is the most simplified ideal situation; the linear distribution model (Model 2) introduces the spatial change of flow velocity; the power function distribution model (Model 3) can simulate more complex actual river situations. By adjusting the parameter $n$, Model 3 can approximate various intermediate situations. Some expressions of flow velocity can be derived based on entropy theory, which are used to describe the relationship between the energy distribution and velocity distribution of water flow in a cross-section [5]. Taking entropy theory as a tool for solving velocity distribution, the progressive model framework of this study demonstrates better scenario adaptability: Model 1 corresponds to the ideal assumption of a uniform flow field, Model 2 can simulate the linear flow velocity gradient of natural river channels, and Model 3, through the adjustment of the power function parameter n, can cover complex distributions from parabolas to high-order non-linearities.

## 3.2 General Solution Framework

The optimization framework used in this study is essentially the engineering application of the calculus of variations. It constructs an objective function (including constraint conditions) through the Lagrange multiplier method, and finally solves the optimal control variable $\theta(y)$ by the calculus of variations. This method not only retains the advantage of the calculus of variations in dealing with the optimization of continuous systems, but also simplifies the mathematical expression of constraint conditions through the multiplier method, providing a unified solution idea for all models: first, establish constraint conditions as $\int_0^L v_x(y)dy/v_y(y) = 0$; second, construct the objective function as $T = \int_0^L d\,y/v_y(y)$; third, apply the calculus of variations to solve the optimal control $\theta(y)$.

## 3.3 Convergence of Numerical Solutions

The feasibility of obtaining valid solutions relies on the parameter $k$ , and the criteria vary across models:

- Model 1 has a feasible solution when $k < 1$
- Model 2 has a feasible solution when $k < 2$
- Model 3 has a feasible solution when $k < \frac{2n+1}{2n}$.

This threshold reflects the matching relationship between the flow velocity gradient intensity ($k$) and the flow field complexity ($n$)—when the spatial change rate of flow velocity exceeds this limit, even the optimal angle control cannot offset the water flow thrust. This is consistent with the practical experience in actual rivers that “ferries in rapid sections need to wait for the water flow to calm down”, and it provides an important design criterion for practical applications.

## 3.4 Analysis of Model Applicability

- Model 1 is suitable for artificial channels or small rivers with relatively uniform flow

velocity

- Model 2 is suitable for natural rivers with obvious central main channels
- Model 3 can describe large river systems with complex cross-sections

### 3.5 Research Limitations

This study focuses on spatially inhomogeneous flow velocities and does not consider time - varying pulsating flow velocities (such as the $v(y,t)$ fluctuations caused by turbulence). For future work, stochastic differential equations can be introduced to establish a "space - time" dual - variable flow velocity model to improve the adaptability to actual rivers. In addition, the convergence analysis of numerical solutions is only based on theoretical derivation, and the universality of the critical value needs to be verified through physical model experiments in the future.

## 4. Conclusion

This study addresses the dynamic problem of a boat crossing a river under variable flow velocities by constructing three typical flow velocity distribution models (constant flow, linear distribution, and even-power function distribution). Employing vector decomposition, calculus, and the Lagrange multiplier method, it derives the analytical expressions of the boat's motion trajectory when the heading relative to the flow is fixed, and solves for the optimal heading angle to reach the direct opposite bank as well as the minimum crossing time. Its core innovation lies in establishing a "simple-to-complex" stepped modeling framework: Model 1 serves as a benchmark for ideal uniform flow, Model 2 introduces linear flow velocity gradients to simulate natural river characteristics, and Model 3 enables the approximation of complex flow fields by adjusting the power parameter n, effectively overcoming the limitations of traditional uniform flow assumptions. Theoretically, this work enriches the teaching and research of kinematics by linking classical physics problems with modern optimization theories; practically, it provides theoretical support for inland ship intelligent navigation (e.g., real-time optimal heading calculation to reduce crossing time and fuel consumption), ferry operation optimization, and emergency rescue path planning, while Model 3 can be extended to analyze ship navigation under the influence of ocean tides. Future research directions include integrating a hull dynamic model to establish a "flow field-ship speed-thrust" coupling equation, optimizing the parameter identification efficiency of Model 3 via machine learning, and verifying the $k$ threshold through physical experiments to further enhance the model's practical applicability. This study demonstrates the practical value of classical physics problems in modern technology and offers a paradigm for bridging theoretical physics and engineering applications.